\documentstyle[aps]{revtex}
\begin{document} 
\draft 
\preprint{gr-qc/9906027}
\title{Parity Violating Gravitational Coupling of Electromagnetic Fields}
\author{Parthasarathi Majumdar$^{\dagger}$\footnote{email:
partha@imsc.ernet.in} and Soumitra 
SenGupta$^{\ddagger}$\footnote{soumitra@juphys.ernet.in}}
\address{$^{\dagger}$ The Institute of Mathematical Sciences, Chennai 600
113, India \\ $^{\ddagger}$ Physics Department, Jadavpur University,
Calcutta 700 032, India.} 
\maketitle 
\begin{abstract}
A manifestly gauge invariant formulation of the coupling of the Maxwell theory
with an Einstein Cartan geometry is given, where the space time torsion
originates from a massless Kalb-Ramond field augmented by suitable $ U(1) $
Chern Simons terms. We focus on the situation where the torsion violates parity,
and relate it to earlier proposals for gravitational parity violation.
\end{abstract}

\section{Introduction}

In Einstein-Maxwell theory, it is well-known that the electromagnetic
field-strength $F_{\mu \nu}$, defined as the generally-covariant curl of the
four-potential $A_{\mu}$, reduces to the flat space expression on account of the
symmetric nature of the Christoffel connection \cite{wald}. However, when
gravitation is taken to be described instead by Einstein-Cartan theory, i.e., a
theory where the connection has an antisymmetric piece (known as spacetime
torsion), the situation changes quite drastically, because the electromagnetic
field strength, again defined through the covariant curl, is no longer gauge
invariant \cite{Hehl}. As we shall argue later, the torsion tensor must
necessarily obey this invariance and as such cannot be used to compensate for the
loss incurred in the field strength tensor. Since electric and magnetic fields are
measurable quantities irrespective of the concomitant existence of a curved
background geometry, this breakdown of gauge invariance is not acceptable.  Thus,
one is led to infer that a coupling of Maxwell electromagnetism to Einstein-Cartan
theory through the minimal coupling prescription is not possible. Attempts have
been made to go beyond the minimal coupling prescription \cite{ni} - \cite{car} by
coupling a certain gauge non-invariant antisymmetric three-tensor constructed out
of the electromagnetic gauge potential (now known as the Chern Simons 3-form) to a
of torsion which is the ({\it Hodge}-) dual of the derivative of a scalar field. 
However, this procedure violates the Einstein Equivalence
Principle if, as in \cite{ni}, the scalar field is stipulated to be a function of
the gravitational field alone. Alternatively, the origin of the scalar field
remains unclear.

Recently, one of us (SS), in collaboration with Mukhopadhyaya \cite{ms} have
explored the possibility of inducing gravitational parity violation through
incorporation of such violation in the torsion tensor itself. In this manner, one
obtains parity violating actions for pure gravity as well as for a host of matter
couplings. However, in that paper, the coupling of gravity to non-gravitational
fields is achieved through the minimal coupling prescription, and, as such, for
reasons mentioned above, cannot be generalized to include electromagnetism. One of
the arenas for observable gravitational parity violation might be the anisotropies
of the Cosmic Microwave Background Radiation (CMBR) \cite{kam}, induced by parity
violating gravito-electromagnetic interactions at a more fundamental level . 
Therefore, it is of utmost importance to ascertain how the Maxwell field couples
to the Einstein-Cartan system, possibly exhibits parity violation in the CMBR
anisotropies, and yet manifestly remains gauge invariant.  

There do exist phenomenological models leading to observable gravitational parity
violations \cite{hari}. Indeed, the spin precession of neutrons in a gravitational
field can lead to parity violating effects which may soon become observable
through NMR studies \cite{harip}. Unfortunately, such phenomenological models do
not ensue from the Einstein Cartan theory; in fact, low energy theorems based
on one-graviton exchange \cite{hari} seem to imply that they are absent in any
theory of gravity based on a {\it symmetric} metric tensor. Apparently, such
parity violations require a {\it propagating} torsion which the Einstein Cartan
theory does not have. However, since these conclusions emerge only in the weak
field approximation, they may not preclude conclusively the possibility of
gravitational parity violation in Einstein Cartan theory. Admittedly,
effects arising from the latter may have somewhat less of a prospect of detection
in laboratory experiments (or even in astrophysical data)  than those
predicted phenomenologically \cite{hari}. On the other hand, they may be
significant in the context of cosmology, as mentioned above.

In this paper, the problem of gauge invariant coupling of the Einstein Cartan
theory to the Maxwell field is reconsidered in the light of wisdom gleaned from
string theory. We focus on a situation where the source of torsion - the
Kalb-Ramond antisymmetric tensor field - violates spatial parity. 

\section{Gauge Invariance and Torsion}

As already stated, the electromagnetic field strength is defined by 
\begin{equation}
F_{\mu\nu} = D_{\mu}A_{\nu} - D_{\nu}A_{\mu}   \label{fmn}
\end{equation}
where the covariant derivative is to be written as 
\begin{equation}
D_{\mu}A_{\nu} = {\partial}_{\mu}A_{\nu} + \Gamma^{\rho}_{\mu\nu} A_{\rho} ~,
\label{cfmn}
\end{equation}
where the affine connection $\Gamma$ includes torsion. Thus the expression for
$F_{\mu\nu}$ in our case turns out to be
\begin{equation}
F_{\mu\nu}~ =~ \partial_{[\mu}A_{\nu]} ~-~2T^{\rho}_{\mu\nu}A_{\rho} \label{gni}
\end{equation}
which is obviously not invariant under the standard $U(1)$ electromagnetic gauge
transformation $\delta A_{\mu} = \partial_{\mu} \omega$. Clearly, this is
quite unacceptable, as the field strengths are measurable quantities even in
a curved spacetime with torsion. Now, the torsion tensor
$T^{\rho}_{\mu \nu}$ - a purely geometric quantity like curvature must be gauge
invariant. To see why this must be so, consider the behaviour of a charged scalar
field under transport along a closed curve, as essentially given by
\begin{equation} 
[~\nabla_{\mu}~,~\nabla_{\nu}~]~\phi=~T^{\rho}_{\mu
\nu}~\nabla_{\rho}~\phi~+~F_{\mu \nu}~\phi~,  \label{comm}
\end{equation}
where, $\nabla_{\mu}~\equiv~D_{\mu} + A_{\mu}$. In (\ref{comm}), the {\it lhs}
transforms exactly like $\phi$ under $U(1)$ gauge transformations, as does the
second term on the {\it rhs} as $F_{\mu \nu}$ is expected to be gauge
invariant. It follows that
the torsion tensor in the first term on the {\it rhs} must therefore be gauge
invariant. Thus, as already mentioned earlier, it is not possible to use it to
compensate for the loss of gauge invariance manifest in (\ref{gni}). 

If indeed no other non-gravitational fields are available, a gauge
invariant coupling of the Maxwell field to torsion might well nigh be
impossible. The situation is far more hopeful if there exists yet another
non-gravitational field, possibly massless, to function as the source of
the torsion. Within the option of bosonic fields, the Kalb-Ramond (KR)
antisymmetric second rank tensor field $B_{\mu \nu}$ appears as a possible
candidate. Indeed, a contact coupling between such a field with the
electromagnetic field strength tensor, has already been proposed
\cite{ham}. Now, since $B_{\mu \nu}$ is a massless antisymmetric field, it
is expected to be a gauge connection, as indeed it is, with the following
gauge transformation $\delta B_{\mu \nu} = \partial_{[\mu} \Lambda_{\nu]}$
which leaves its field strength $H_{\mu \nu
\lambda}~\equiv~\partial_{[\mu} B_{\nu \lambda]}$ gauge invariant.
Unfortunately, in \cite{ham} (and other earlier works on this theme), this
invariance has not been sufficiently respected.

There is yet another reason to consider the massless KR field: it is an
inescapable part of the massless spectrum of any critical string theory,
appearing upon compactification to standard four dimensional spacetime
\cite{gsw}. The gauge invariance mentioned above is of course well-known
in the string context.

\section{Gauge Invariant Einstein-Cartan-Maxwell-Kalb-Ramond Coupling}

The first step is to realize that in order to obtain a coupling that is
invariant under both electromagnetic and Kalb-Ramond gauge
transformations, the KR tensor potential must be endowed with a
non-trivial electromagnetic gauge transformation property, and the KR
field strength must be modified with the addition of an electromagnetic
Chern Simons three tensor. One of the motivations of the latter comes from
string theory \cite{gsw}, where the KR 3-form $H$ is modified by addition of
Yang-Mills and gravitational Chern Simons 3-forms to ensure that the
quantum theory is anomaly free. Thus, the modified KR field strength
three-tensor in our case is defined as
\begin{equation}
{\tilde H}_{\mu \nu\lambda}~\equiv~  H_{\mu \nu\lambda}~+~\frac13~A_{[\mu}~F_{\nu
\lambda]}~. \label{krh}
\end{equation}
This modified tensor ${\tilde H}_{\mu \nu\lambda}$ is gauge invariant under standard
$U(1)$ gauge transformations, provided we stipulate that the KR potential transforms
under $U(1)$ gauge transformations as $\delta B_{\mu \nu} = -\omega~F_{\mu \nu}$. We
have of course used the standard Bianchi identities for the Maxwell field involving
the Christoffel connection. Further, the Maxwell field is assumed to be left
invariant under KR gauge transformations.

We now propose the following action for a manifestly gauge invariant
Einstein-Cartan-Maxwell-Kalb-Ramond coupling,
\begin{eqnarray}
S ~&=&~ \int d^{4}x~ \sqrt{-g}~[~ R~(g~,~T)~-~\frac14 F_{\mu \nu} ~F^{\mu
\nu}~-\frac12~{\tilde H}_{\mu \nu \lambda} {\tilde H}^{\mu \nu \lambda}~\nonumber \\
&+&~T^{\mu \nu \lambda}~{\tilde H}_{\mu \nu \lambda}~ ]~\label{act}
\end{eqnarray}
where $R$ is the scalar curvature, defined as 
$R = R_{\alpha\mu\beta\nu}g^{\alpha\beta}g^{\mu\nu}$. $R_{\alpha\mu\beta\nu}$ 
is the Riemann-Christoffel tensor:
\begin{equation}
R^{\kappa}_{\mu\nu\lambda} =
\partial_{\mu} \Gamma^{\kappa}_{{\nu\lambda}}
- \partial_{\nu} \Gamma^{\kappa}_{{\mu\lambda}} + 
\Gamma^{\kappa}_{\mu\sigma} \Gamma^{\sigma}_{\nu\lambda} - 
 \Gamma^{\kappa}_{\nu\sigma} \Gamma^{\sigma}_{\mu\lambda}  
\end{equation}    

The torsion tensor $T_{\mu \nu \lambda}$ is an auxiliary field in eq.
(\ref{act}),
obeying the constraint equation
\begin{equation}
T_{\mu \nu \lambda}~~=~~{\tilde H}_{\mu \nu \lambda}~. \label{tors}
\end{equation}
Thus, the augmented KR field strength three tensor plays the role of the
spin angular momentum density \cite{Hehl}. Substituting the above equation
in (\ref{act}) and varying with respect to $B_{\mu \nu}$ and $A_{\mu}$
respectively, we obtain the equations
\begin{equation} 
D^S_{\mu}~{\tilde H}^{\mu \nu \lambda}~~=~~0 \label{bmn}
\end{equation}
and
\begin{equation}
D^S_{\mu}~F^{\mu \nu}~~=~~{\tilde H}^{\mu \nu \lambda}~ F_{\lambda \mu}~,
\label{max}
\end{equation}
where, $D^S$ is the covariant derivative using the Christoffel connection.
Clearly, these equations of motion are manifestly gauge covariant under
both gauge transformations. The interaction term thus has the structure 
\begin{equation}
S_{int}~=~\int d^3 x~\sqrt{-g}~ H^{\mu \nu \lambda}~A_{\mu} F_{\nu \lambda}
~.\label{intr}
\end{equation}
Such a structure has been proposed earlier \cite{des} on quite different
grounds to solve the problem of gauge invariant Einstein-Cartan-Maxwell
couplings. Since the KR three tensor is Hodge-dual to the derivative of a
spinless field $\phi$, so that, after a partial integration, one obtains,
\begin{equation}
S_{int}~=~\frac12~\int d^4 x ~\phi~F_{\mu \nu}~^*F^{\mu \nu}~,
\label{inter}
\end{equation}
where, $^*F^{\mu \nu}~\equiv~\epsilon^{\mu \nu \lambda \sigma}F_{\lambda
\sigma}$. Here, we have noted the fact that 
\begin{equation}
{1 \over \sqrt{-g}} \partial_{\nu}~(\sqrt{-g} ~^*F^{\mu
\nu})~=~D^S_{\mu}~^*F^{\mu \nu}~=~0
\end{equation}
by the Maxwell Bianchi identity. 

In standard string theory inspired applications to high energy
phenomenology, the spinless field is considered to be a pseudoscalar -
prospectively an axion. Here we depart from this interpretation. In ref. 
\cite{ms}, it has been shown how to construct parity violating actions in
pure gravity as well as with matter couplings, by starting with a parity
violating torsion tensor and using the minimal coupling prescription. 
Although the above construction of torsion coupling of the electromagnetic
field using the KR antisymmetric tensor field is not minimal in the sense of
\cite{ms}, parity violation can easily be incorporated here as well. 
Indeed, it has already been noted \cite{kam} that the interaction
(\ref{inter}) above induces parity violation for the electromagnetic field,
provided $\phi$ is a {\it scalar} field. However, such a proposal is in need
of a firmer theoretical underpinning which our derivation provides. At the
level of the antisymmetric KR field, the proposal that $\phi$ is a
scalar may be interpreted as stipulating that KR field has the wrong
parity, as in ref. \cite{ms}. Notice that the crucial
ingredient in inducing gauge invariant parity violating couplings of the
electromagnetic field has been the augmentation of the KR field strength by
the Chern Simons term. 

\section{Discussion and Conclusions}

Two issues have been addressed in the foregoing : that of coupling torsion to
the Maxwell field in a way that respects all Abelian gauge symmetries, and that
of using such a coupling to obtain gauge invariant parity violation for the
electromagnetic field. The generic Lagrangian density as depicted in
(\ref{inter}) has been proposed earlier \cite{ni} as a counterexample to the
Einstein Equivalence Principle, where the scalar field, however, is stipulated
to be a function of the gravitational fields alone. The observability of the
consequent rotation in the polarization plane of synchrotron radiation from
distant sources has also been explored earlier \cite{car} and found to be
insignificant. An action similar to, although not identical to, (\ref{inter})
giving a gauge invariant coupling of the Maxwell field to the Einstein Cartan
system has also been considered \cite{des}. However, in the latter work, little
is said about the possible origin of the so-called torsion potential. In our
opinion, our work ties up the loose ends of earlier approaches into one
consistent framework which could in principle lead to observable predictions of
gravitational parity violation. 

In the recent past, some analysis of data on synchrotron radiation from
distant radio galaxies has led to claims \cite{nod}, albeit a bit
controversial \cite{carr}, that the rotation of the plane of polarization of
the radiation has indeed been observed. If these claims are valid, they
could provide a testing ground for our proposals here. Indeed, in a somewhat
related approach \cite{doba}, a massive fermion is coupled to a background
torsion and integrated to produce a one-loop effective action of the type
(\ref{inter}). However, regularization issues pertaining to the axial
anomaly do not seem to have been adequately addressed in this work. If the
theory is rendered completely free of gauge and gravitational anomalies,
such effects may actually even disappear. Furthermore, the background torsion
used in the paper has uncertain origins. In contrast, our augmentation of
the KR field strength is motivated by the requirement of anomaly freedom in
string theory and perhaps is therefore more consistent.

In this context, it may be noted that the augmentation as given in
(\ref{krh}) is actually incomplete, since the Lorentz Chern Simons three form
has not been added.  This is necessitated by the requirement of freedom from
local Lorentz anomalies \cite{gsw}. It is not difficult to show that the
addition of such
terms to the ${\tilde H}$ in (\ref{krh}) would generate additional parity
violating couplings of the type
\begin{equation}
S_{int}~=~\int d^4x~\phi ~\epsilon^{\mu \nu \rho \lambda}~R_{\mu \nu \eta
\sigma}~ R_{\rho \lambda}~^{\eta \sigma}~  \label{rrdul}
\end{equation}
which might show up in the polarization asymmetry of gravitational waves
\cite{kam}.  

Finally we remark that any observation of the effects that we have alluded to
here can be construed to imply the existence of spacetime torsion, albeit 
locally in spacetime, at least in the earliest epoch.

\noindent {\bf Acknowledgements}: One of us (PM) thanks N. D. Hari Dass  and
R. Kaul for illuminating discussions. He also thanks S. Kar for correspondence
and for bringing ref. \cite{doba} to his attention. This work was supported by
Project Grant no. 98/37/16/BRNS cell/676 from the Board of Research in Nuclear
Sciences, Department of Atomic Energy, Government of India.

\end{document}